\documentclass[12pt]{article}
\usepackage{amsmath ,amssymb}
\usepackage{amsthm}
\usepackage{graphicx}

\usepackage{typearea}
\typearea{13}

\newcommand{\set}{\setlength\arraycolsep{2pt}} 

\newcommand{\e}{\epsilon}

\newcommand{\dl}{\delta}

\newcommand{\om}{\omega}
\newcommand{\tom}{\tilde{\omega}}
\newcommand{\Om}{\Omega}
\newcommand{\tOm}{\tilde{\Omega}}
\newcommand{\la}{\lambda}

\newcommand{\Sg}{\Sigma}

\newcommand{\Bcal}{\mathcal{B}}

\newcommand{\Fcal}{\mathcal{F}}

\newcommand{\Kcal}{\mathcal{K}}
\newcommand{\Lcal}{\mathcal{L}}

\newcommand{\Ncal}{\mathcal{N}}

\newcommand{\Vcal}{\mathcal{V}}

\newcommand{\Zcal}{\mathcal{Z}}

\newcommand{\lp}{\left(}
\newcommand{\rp}{\right)}
\newcommand{\lc}{\left\{}

\newcommand{\lb}{\left[}
\newcommand{\rb}{\right]}

\newcommand{\ra}{\rightarrow}

\newcommand{\RA}{\Rightarrow}
\newcommand{\p}{\partial}
\newcommand{\w}{\!\wedge\!}

\newcommand{\Tr}{\text{Tr}}

\newcommand{\non}{\nonumber} 
\newcommand{\ba}{\begin{alignat}{3}}
\newcommand{\ea}{\end{alignat}}

\newcommand{\WLR}{\langle W_{R} \rangle} 
\newcommand{\WLA}{\langle W_{A_{k}} \rangle} 
\newcommand{\WLS}{\langle W_{S_{k}} \rangle} 

\newcommand{\lpl}{\ell_{\mathrm{P}}}
\DeclareMathOperator{\sign}{\mathrm{sign}}
\newcommand{\AdS}{\mathrm{AdS}}

\numberwithin{equation}{section}

\allowdisplaybreaks

\begin{document}

\renewcommand{\thefootnote}{\fnsymbol{footnote}}
\setcounter{page}{0}
\thispagestyle{empty}
\begin{flushright}
OU-HET 810 
\\
\end{flushright}

\vskip3cm
\begin{center}
{\LARGE {\bf M5-branes and Wilson Surfaces in AdS$_{7}$/CFT$_{6}$ Correspondence}} 
\vskip1cm
{\large 
{\bf Hironori Mori\footnote{\href{mailto:hiromori@het.phys.sci.osaka-u.ac.jp}{hiromori@het.phys.sci.osaka-u.ac.jp}}
and
Satoshi Yamaguchi\footnote{\href{mailto:yamaguch@het.phys.sci.osaka-u.ac.jp}{yamaguch@het.phys.sci.osaka-u.ac.jp}}} 

\vskip1cm
\it Department of Physics, Graduate School of Science, 
\\
Osaka University, Toyonaka, Osaka 560-0043, Japan}
\end{center}

\vskip1cm
\begin{abstract} 
We study $\AdS_{7}$/CFT$_{6}$ correspondence between M-theory on $\AdS_{7} \times S^{4}$ and the 6D $\Ncal = (2,0)$ superconformal field theory. In particular we focus on Wilson surfaces. We use the conjecture that the (2,0) theory compactified on $S^{1}$ is equivalent to the 5D maximal super Yang-Mills (MSYM) and Wilson surfaces wrapping this $S^1$ correspond to Wilson loops in 5D MSYM.  The Wilson loops in 5D MSYM obtained by the localization technique result in the Chern-Simons matrix model. We calculate the expectation values of Wilson surfaces in large-rank symmetric representations and anti-symmetric representations by using this result. On the gravity side, the expectation values for probe M5-branes wrapping submanifolds of the background are computed. Consequently we find new, nontrivial evidence for the $\AdS_{7}$/CFT$_{6}$ correspondence that the results on the gravity side perfectly agree with those on the CFT side.
\end{abstract}

\renewcommand{\thefootnote}{\arabic{footnote}}
\setcounter{footnote}{0}

\vfill\eject

\tableofcontents

\section{Introduction and summary} \label{Intro}
The AdS/CFT correspondence \cite{Maldacena:1997re} provides a lot of new insight into wide regions of physics, and it is significant to reveal even more properties of this duality for understanding string theories and gauge theories. While many attempts succeed in confirming it in lower dimensions, the higher-dimensional versions of the correspondence are still mysterious. The main reason is that there are few known facts about conformal field theories in higher dimensions. However, recently it was found that the supersymmetric localization can be applied to 5D super Yang-Mills theories on curved geometries and their partition functions can be derived exactly as mentioned below. We can utilize them to verify the $\AdS_{d + 1}$/CFT$_{d}$ for $d \geq 5$. For example, there are a few pieces of evidence of the $\AdS_{6}$/CFT$_{5}$ \cite{Jafferis:2012iv,Assel:2012nf}.

The 5D $\Ncal = 1$ super Yang-Mills theories are constructed on several curved backgrounds.  Their partition functions and expectation values of Wilson loops have been calculated by the localization technique \cite{Kallen:2012cs, Hosomichi:2012ek, Kallen:2012va, Kim:2012ava, Imamura:2012xg, Kim:2012tr, Imamura:2012bm, Kim:2012qf, Qiu:2013pta, Kim:2013nva, Qiu:2013aga, Schmude:2014lfa}. 
The 5D $\Ncal=1^{*}$ theory on the round five-sphere with a radius $r$, which contains a vector multiplet and an adjoint hypermultiplet, has $\Ncal=2$ supersymmetry if the mass for the hypermultiplet takes a specific value.
Then the partition functions and Wilson loops reduce to the Chern-Simons matrix model \cite{Kim:2012ava, Minahan:2013jwa} first considered in \cite{Marino:2002fk}. Also, we can produce the 5D maximal super Yang-Mills (MSYM) on $S^{5}$ from the (2,0) theory by the dimensional reduction with the appropriate twist to keep the supersymmetry \cite{Kim:2012ava}. It is argued in \cite{Witten:1995ex, Douglas:2010iu, Lambert:2010iw} that Kaluza-Klein modes in 6D can be identified with instanton particles in 5D under
\begin{align}
R_{6} = \frac{g_{YM}^{2}}{8 \pi^{2}}, \label{KKinst}
\end{align}
where $R_{6}$ is the radius of the compactified $S^{1}$, and $g_{YM}$ is the five-dimensional gauge coupling constant. Following their discussion, the 5D MSYM seems to contain all degrees of freedom of the (2,0) theory. An observation supporting this claim is that the free energy obtained by the Chern-Simons matrix model reproduces $N^3$ behavior\footnote{However, the free energies for the Chern-Simons matrix model and the supergravity do not completely coincide by an overall constant \cite{Kallen:2012zn, Minahan:2013jwa}.} of the supergravity analysis on $\AdS_{7} \times S^{4}$ \cite{Kim:2012ava, Minahan:2013jwa, Kallen:2012zn, Giasemidis:2013oea}.

In this paper, we focus on the expectation values of Wilson surfaces for the $\AdS_{7}$/CFT$_{6}$ correspondence. The Wilson surfaces in the (2,0) theory are a class of nonlocal operators localized on surfaces in 6D \cite{Ganor:1996nf}. Through the above argument, Wilson surfaces extending to the compatified direction are Wilson loops in the 5D theory. Therefore, we compute the expectation values of them by using the Chern-Simons matrix model.  In particular we evaluate the expectation values of Wilson loops in large-rank anti-symmetric representations and symmetric representations in the large $N$ limit.

On the other hand, naively a probe M2-brane ending on multiple M5-branes is the M-theory description of the Wilson surface \cite{Ganor:1996nf}.  The holographical description of a spherical Wilson surface has been studied in \cite{Berenstein:1998ij,Corrado:1999pi}. Recently, it has been clarified in \cite{Young:2011aa, Kim:2012qf, Minahan:2013jwa} that the expectation value of the Wilson surface wrapping on $S^1\times S^1$ in the fundamental representation matches that of the M2-brane wrapping $\AdS_{3}$. 

In this paper, we consider a probe M5-brane description of the Wilson surface \cite{Lunin:2007ab,Chen:2007ir,Chen:2007zzr,Chen:2008ds,D'Hoker:2008qm} instead of the M2-brane.  When the number of the overlapping and winding M2-branes becomes large, they blow up and make an M5-brane with worldvolume flux wrapping two types of submanifolds of $\AdS_{7} \times S^{4}$ due to the representation: one is $\AdS_{3} \times S^{3}$ totally in $\AdS_{7}$, and the other is $\AdS_{3} \times \tilde{S}^{3}$ belonging to $S^{4}$.  This is the analogue of the D3-brane and D5-brane description of the symmetric and anti-symmetric Wilson loops in $\AdS_5$/CFT$_4$ correspondence\cite{Rey:1998ik,Drukker:2005kx,Hartnoll:2006hr,Yamaguchi:2006tq,Gomis:2006sb}.  According to this analogy, we expect that an M5-brane wrapping on $\AdS_{3} \times S^{3}$ corresponds to the symmetric representation and one wrapping on $\AdS_{3} \times \tilde{S}^{3}$ corresponds to the anti-symmetric representation.  We calculate the expectation values of the Wilson surfaces by evaluating the on-shell action of these M5-branes. In the calculation for the M5-branes, we use the so-called Pasti-Sorokin-Tonin (PST) action \cite{Pasti:1997gx, Bandos:1997ui, Bandos:1997gm}.

We compare the results on the CFT side and the gravity side, and we find new evidence supporting the $\AdS_{7}$/CFT$_{6}$ correspondence; the M5-brane wrapping $\AdS_{3} \times S^{3}$ and wrapping $\AdS_{3} \times \tilde{S} ^{3}$ perfectly agree with the Wilson surface in symmetric representation and in anti-symmetric representation respectively. We note that the authors of \cite{Minahan:2013jwa} have suggested that the relation \eqref{KKinst} be modified at strong coupling such that the constant coefficient becomes dependent on the square of the mass for the adjoint hypermultiplet. One can find that our results are truly consistent with their argument.

One of the interesting future directions is to study the relation between the bubbling geometry and Wilson surfaces in larger representations.
A class of bubbling solutions in the 11-dimensional supergravity as the gravity dual of the Wilson surfaces is obtained in \cite{Yamaguchi:2006te,Lunin:2007ab,D'Hoker:2008wc,D'Hoker:2008qm} along the line of the bubbling geometry for local operators \cite{Lin:2004nb} and Wilson loops \cite{Yamaguchi:2006te,Lunin:2006xr,D'Hoker:2007fq}.  In these solutions, the eigenvalue distribution of the matrix model is suggested as the following form: the real line of the eigenvalue space is divided into black and white segments, and the density is a positive constant on the black segments and zero on the white segments. The unit length of a black segment is twice that of a white segment. Actually, the eigenvalue distribution of the Chern-Simons matrix model obtained in \cite{Halmagyi:2007rw} is consistent with the bubbling solutions.  This observation is other evidence of the correspondence.  It will be an interesting future work to calculate the expectation values of Wilson surfaces by using the bubbling solutions and compare them to the calculation in the Chern-Simons matrix model.

The rest of the paper is organized as follows: In section \ref{CSMM} we use the Chern-Simons matrix model and evaluate the expectation values of Wilson surfaces in anti-symmetric representation and symmetric representation.
In section \ref{M5}, we use probe M5-branes on the gravity side and calculate the expectation values of the Wilson surfaces.

\section{Wilson surfaces in Chern-Simons matrix model} \label{CSMM}
\subsection{Chern-Simons matrix model in large $N$}
We consider 6D A$_{N-1}$ type (2,0) theory on $S^1\times S^5$ and a Wilson surface in this theory.   This Wilson surface is wrapping on $S^1\times S^1$ where the first $S^1$ is orthogonal to $S^5$ and the second $S^1$ is a great circle of $S^5$. This Wilson surface can be treated as a Wilson loop wrapping a great circle in 5D SU$(N)$ MSYM on $S^5$ if the boundary condition in the $S^1$ direction is twisted appropriately \cite{Kim:2012ava,Kim:2012qf}.

The expectation values of Wilson loops wrapping on the great circle on $S^5$ with a radius $r$ are calculated by using the localization technique \cite{Kallen:2012cs, Hosomichi:2012ek, Kallen:2012va, Kim:2012ava, Kim:2012qf}.  In particular, the expectation value of the Wilson loop in the representation $R$ in MSYM with a coupling constant $g_{YM}$ reduces to the Chern-Simons matrix model
\begin{align}
\WLR
&= \frac{1}{\mathcal{Z}} \int \prod_{i = 1}^{N} d \nu_{i} \prod_{i, j, i \neq j} \left|\sinh \frac{N}{2}(\nu_{i} - \nu_{j}) \right|
\exp \lb - \frac{N^{2}}{\beta} \sum_{i = 1}^{N} \nu_{i}^{2} \rb \Tr_{R} e^{N \nu}, \label{WLR}
\end{align}
where $\beta = \frac{g_{YM}^{2}}{2 \pi r}$.
$\Zcal$ is the partition function given by
\begin{align}
\Zcal
&:= \int \prod_{i = 1}^{N} d \nu_{i} \prod_{i, j, i \neq j} \left|\sinh \frac{N}{2}(\nu_{i} - \nu_{j}) \right|
\exp \lb - \frac{N^{2}}{\beta} \sum_{i = 1}^{N} \nu_{i}^{2} \rb. \label{CSMMpf}
\end{align}
We evaluate these integrals in the limit $N\to\infty$ while $\beta$ is kept finite in order to compare them to the gravity calculation. Notice that this limit is different from the 't Hooft limit. For the 't Hooft limit, the expectation values of the Wilson loops are computed in \cite{Drukker:2010nc}.

Let us first consider the eigenvalue distribution of the partition function before calculating the Wilson loop. When we take $N \ra \infty$ with fixed $\beta$, the hyperbolic sine factor is simplified and we obtain
\begin{eqnarray} 
\mathcal{Z} \sim \int \prod_{i = 1}^{N} d \nu_{i} \exp \lb - \frac{N^{2}}{\beta} \sum_{i = 1}^{N} \nu_{i}^{2} + \frac{N}{2} \sum_{i, j, i \neq j} \left| \nu_{i} - \nu_{j} \right| \rb.
\end{eqnarray}
In the limit, both terms in the exponential are $O ( N^{3} )$, and, therefore, this integral can be evaluated by saddle points. It yields to the saddle point equations for $\nu_{i}$
\begin{eqnarray} 
0 = - \frac{2 N^{2}}{\beta} \nu_{i} + N \sum_{j, i \neq j} {\rm sign} (\nu_{i} - \nu_{j} ).
\end{eqnarray}
We can easily find the following solutions under the assumption $\nu_{i} > \nu_{j}$ for $i < j$:
\begin{eqnarray} 
\nu_{i} = \frac{\beta}{2 N} ( N - 2 i ). \label{nui}
\end{eqnarray}
In other words the eigenvalue density is given by
\begin{eqnarray} 
\rho ( \nu ) = \lc \begin{aligned}
\frac{1}{\beta} & \hspace{1em} \mbox{for } | \nu | \leq \frac{\beta}{2}, \\
0 & \hspace{1em} \mbox{for } | \nu | > \frac{\beta}{2}.
\end{aligned} \right.
\end{eqnarray}

We note that instanton factors do not appear in our computation. The full partition function of the $\Ncal = 1$ SYM on $S^{5}$ including instantons is derived in \cite{Kim:2012qf} as
\begin{align}
\Zcal ( \beta, m, \e_{1}, \e_{2} ) \sim \int \prod_{i = 1}^{N} d \nu_{i} \exp \lb - \frac{N^{2}}{\beta ( 1 + a ) ( 1 + b ) ( 1 + c )} \sum_{i = 1}^{N} \nu_{i}^{2} \rb \prod_{A = 1}^{3} \Zcal_{\rm pert}^{( A )} \Zcal_{\rm inst}^{( A )},
\end{align}
where $\Zcal_{\rm inst}^{( A )}$ is an instanton one-loop determinant (see \cite{Kim:2012qf} for details). For the maximally supersymmetric case obtained by taking appropriate limits of each parameter, the perturbative part $\Zcal_{\rm pert}^{( 1 )} \Zcal_{\rm pert}^{( 2 )} \Zcal_{\rm pert}^{( 3 )}$ reduces to \eqref{CSMMpf} and
\begin{align}
\Zcal_{\rm inst}^{( 1 )} \Zcal_{\rm inst}^{( 2 )} \Zcal_{\rm inst}^{( 3 )}
\to e^{\frac{N \pi^{2}}{3 \beta}} \prod_{n = 1}^{\infty} \lp 1 - e^{- \frac{8 \pi^{2} n}{\beta}} \rp^{- N}
= \eta ( e^{- \frac{8 \pi^{2}}{\beta}} )^{- N}.
\end{align}
Thus, the instanton factor in MSYM is just a constant independent of the integration valuables and does not affect the expectation value because this should be canceled by the normalization factor in \eqref{WLR}.


\subsection{Symmetric representation}
Let us consider symmetric representation $S_{k}$ where the rank $k$ is $O(N)$.
The trace in $S_{k}$ is expressed as
\begin{eqnarray} 
\Tr_{S_{k}} e^{N \nu} = \sum_{1 \leq i_{1} \leq \cdots \leq i_{k} \leq N} \exp \lb N \sum_{l = 1}^{k} \nu_{i_{l}} \rb. \label{symmetric}
\end{eqnarray}
Although \eqref{symmetric} includes various contributions in the summation, the largest one comes with $\nu_{1} = \nu_{i_{1}} = \cdots = \nu_{i_{k}}$. Therefore, the leading contribution to the expectation value is given by
\begin{eqnarray} 
\WLS \sim \int \prod_{i = 1}^{N} d \nu_{i} \exp \lb - \frac{N^{2}}{\beta} \sum_{i = 1}^{N} \nu_{i}^{2} + \frac{N}{2} \sum_{i, j, i \neq j} \left| \nu_{i} - \nu_{j} \right| + N k \nu_{1} \rb. \label{WLS}
\end{eqnarray}
We again acquire $\nu_{1}$ by the saddle point equation
\begin{eqnarray} 
0 = - \frac{2 N^{2}}{\beta} \nu_{1} + N \sum_{j=2}^{N} ( + 1 ) + N k.
\end{eqnarray}
Hence,
\begin{eqnarray}
\nu_{1} = \frac{\beta}{2 N} ( N + k ). \label{nu1s}
\end{eqnarray}
We put it back into \eqref{WLS}, then the leading one depending on $k$ becomes
{\set
\begin{eqnarray} 
\WLS
&\sim& \left. \exp \lb - \frac{N^{2}}{\beta} \sum_{i = 1}^{N} \nu_{i}^{2} + \frac{N}{2} \sum_{i, j, i \neq j} \left| \nu_{i} - \nu_{j} \right| + N k \nu_{1} \rb \right|_{\rm saddle\ point} \non \\
&\sim& \left. \exp \lb - \frac{N^{2}}{\beta} \nu_{1}^{2} + N \sum_{j = 2}^{N} \left| \nu_{1} - \nu_{j} \right| + N k \nu_{1} + ( \mbox{terms independent of } k ) \rb \right|_{\rm saddle\ point} \non \\
&\sim& \exp \lb \frac{\beta}{2} N k \left(1 + \frac{k}{2N} \right) \rb. \label{WLs}
\end{eqnarray} }%
Here we use the fact that $\WLS=1$ when $k=0$. 
This expression \eqref{WLs} reproduces the result of the fundamental case when $k  =1$ \cite{Kim:2012qf, Minahan:2013jwa}. The same result as \eqref{WLs} is also obtained by substituting $n=1,\ m=k$ in \eqref{WRect} or \eqref{WRectSU}.  This result \eqref{WLs} is compared to the result on the gravity side in the next section.

\subsection{anti-symmetric representation}
We turn to calculating the expectation value of the Wilson loop in anti-symmetric representation $A_{k}$ with $k=O(N)$ boxes in the Young diagram. The trace in this representation is written as
\begin{eqnarray} 
\Tr_{A_{k}} e^{N \nu} = \sum_{1 \leq i_{1} < \cdots < i_{k} \leq N} \exp \lb N \sum_{l = 1}^{k} \nu_{i_{l}} \rb.
\end{eqnarray}
The largest contribution in the large $N$ limit is in the case of $i_{l} = l$ because of our ordering $\nu_{1} > \nu_{2} > \cdots > \nu_{N}$, namely, the leading one in \eqref{WLR} is
\begin{eqnarray}
\WLA \sim \int \prod_{i = 1}^{N} d \nu_{i} \exp \lb - \frac{N^{2}}{\beta} \sum_{i = 1}^{N} \nu_{i}^{2} + \frac{N}{2} \sum_{i, j, i \neq j} \left| \nu_{i} - \nu_{j} \right| + N \sum_{l = 1}^{k} \nu_{l} \rb. \label{WLA}
\end{eqnarray}
Since this insertion does not change the eigenvalue distribution, we can find with \eqref{nui},
{\set
\begin{eqnarray} 
\WLA
&\sim& \left. \exp \lb N \sum_{l = 1}^{k} \nu_{l} \rb \right|_{\rm saddle\ point} \non \\ 
&\sim& \exp \lb \frac{\beta}{2} Nk \left( 1 - \frac{k}{N} \right) \rb. \label{WLa} 
\end{eqnarray} }The expression is invariant under the exchange of $k$ and $(N - k)$ as expected and reproduces the result of the fundamental case when $k  =1$ \cite{Kim:2012qf, Minahan:2013jwa}.
The same result as \eqref{WLa} is also obtained by substituting $n=k,\ m=1$ in \eqref{WRect} or \eqref{WRectSU}.  This result \eqref{WLa} is compared to the result on the gravity side in the next section.

\section{Probe M5-branes in 11D supergravity} \label{M5}
Let us now turn to the holographic description of the Wilson surfaces. An M2-brane wrapping $\AdS_{3}$ is the gravity dual to the Wilson surface in fundamental representation \cite{Berenstein:1998ij, Corrado:1999pi, Young:2011aa, Kim:2012qf, Minahan:2013jwa}.  On the other hand,  probe M5-branes are better descriptions for the Wilson loops in large-rank symmetric or anti-symmetric representations \cite{Lunin:2007ab,Chen:2007ir,Chen:2007zzr,Chen:2008ds,D'Hoker:2008qm}, and we employ this probe M5-brane description in this paper. 

\subsection{Supergravity background}
\begin{figure}[t] 
	\begin{center}
	\includegraphics[width=9cm,trim=50 80 50 80,clip]{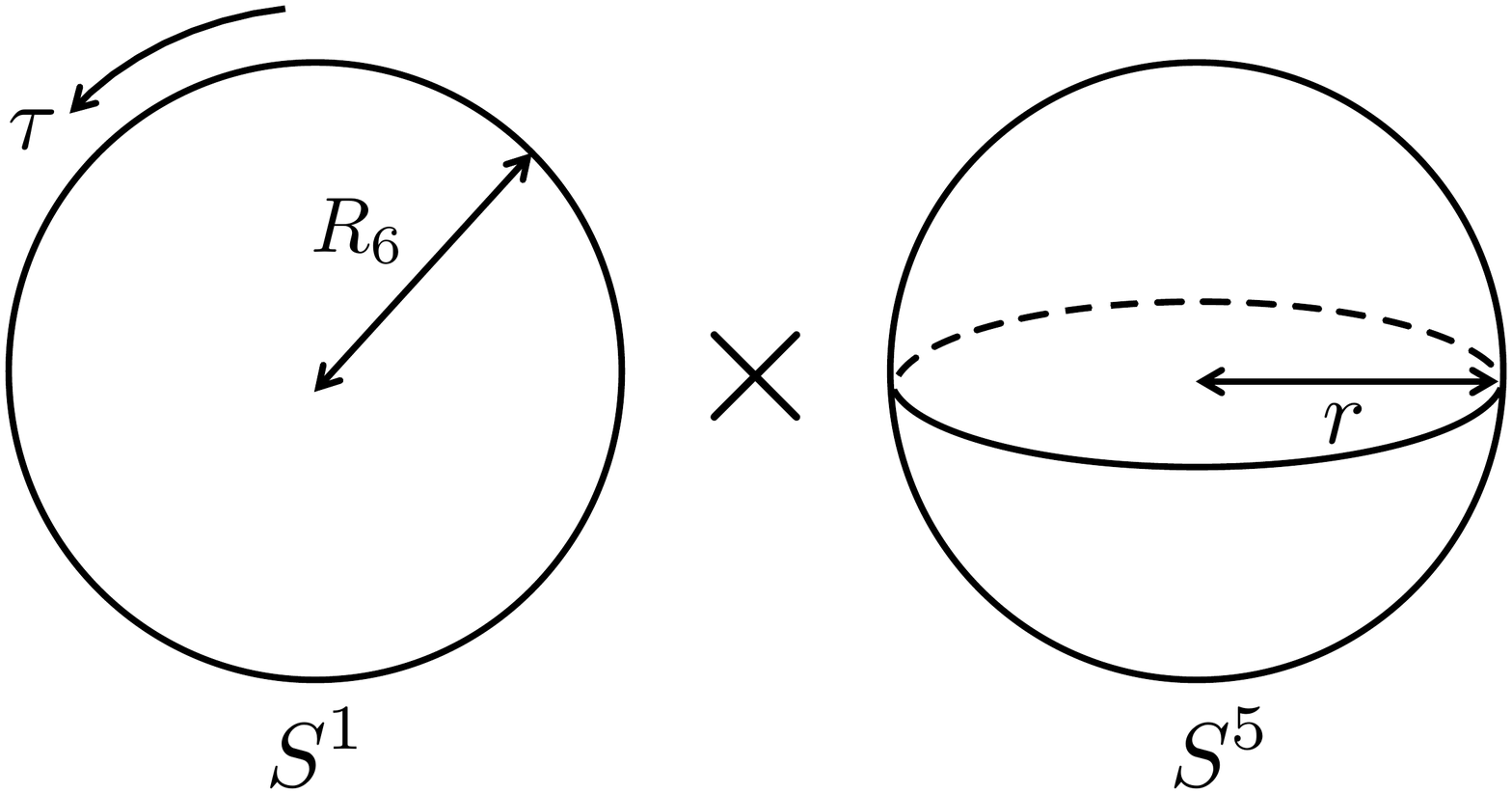}
	\caption{The boundary of $\AdS_{7}$ in the global coordinates. The radii of $S^{1}$ and $S^{5}$ on the boundary are $R_{6}$ and $r$, respectively.}
	\label{bdy1}
	\end{center}
\end{figure}

We take the following forms for the AdS radius $L$ and the M5-brane tension $T_{5}$ as well as in \cite{Maldacena:1997re}
\begin{eqnarray} 
L = 2 \lp \pi N \rp^{\frac{1}{3}} \lpl, \hspace{2em}
T_{5} = \frac{1}{( 2 \pi )^{5} \lpl^{6}},
\end{eqnarray}
where $\lpl$ is the 11-dimensional Planck length.
The metric of Euclidean $\AdS_7\times S^4$ is written in terms of the global coordinates
\begin{equation} 
	\begin{aligned}
	d s^{2} &= L^{2} \lp \cosh^{2} \rho d \tau^{2} + d \rho^{2} + \sinh^{2} \rho d \Om_{5}^{2} \rp + \frac{L^{2}}{4} d \Om_{4}^{2}, \\ 
	d \Om_{5}^{2} &= d \eta^{2} + \sin^{2} \eta d \phi^{2} + \cos^{2} \eta d \Om_{3}^{2}, \\ 
	d \Om_{4}^{2} &= d \theta^{2} + \sin^{2} \theta d \tOm_{3}^{2}, \\
	&\rho\ge 0,\quad 0\le\phi<2\pi,\quad 0\le\eta\le \frac{\pi}{2},\quad 0\le \theta \le \pi,
	\end{aligned} \label{global}
\end{equation}
where $d \Om_{3}^{2}$ and $d \tOm_{3}^{2}$ are metrics of units $S^3$ and $\tilde{S}^3$, respectively.  In order to make the boundary $S^1\times S^5$, we compactify the $\tau$ direction (see Fig. \ref{bdy1}) as
\begin{align}
\tau \sim \tau + \frac{2 \pi R_{6}}{r}.  \label{identification}
\end{align}
To be precise, the identification \eqref{identification} is accompanied by the rotation of the isometry in the $\tilde{S}^3$ direction in order to compare the result from 5D MSYM \cite{Kim:2012ava,Kim:2012qf}. 

Another convenient set of coordinates is the $\AdS_3\times S^3$ foliation.  In these coordinates, the metric is expressed as
\begin{eqnarray} 
	\begin{aligned}
	d s^{2} &= L^{2} \lp \cosh^{2} u d \check{\Om}_{3}^{2} + d u^{2} + \sinh^{2} u d \Om_{3}^{2} \rp + \frac{L^{2}}{4} d \Om_{4}^{2}, \\ 
	d \check{\Om}_{3}^{2} &= \cosh^{2} w d \tau^{2} + d w^{2} + \sinh^{2} w d \phi^{2}, 
	\end{aligned} \label{bubbling}
\end{eqnarray}
where $(u,w)$ are related to $(\rho,\eta)$ as
\begin{align}
&\sinh u=\sinh\rho\cos\eta,\\
&\tanh w=\tanh \rho \sin \eta.
\end{align}
We denote the vielbein for the spacetime by $E^{a}$, then divide each component such as $( E^{0}, E^{1}, E^{2} )$ for $\AdS_{3}$, $E^3=Ldu$, $( E^{4}, E^{5}, E^{6} )$ for $S^{3}$ belonging to $\AdS_{7}$, $E^7=Ld\theta$, and $(E^{8}, E^{9}, E^{\natural} )$ ($\natural = 10$) for $\tilde {S}^{3}$ in $S^4$. 

The supergravity in 11 dimensions contains the 4-form field strength $B_{4}$ as a bosonic field besides the metric.  When the background geometry is $\AdS_{7} \times S^{4}$, 4-form field strength $B_{4}$ is given by
\begin{eqnarray}
B_{4} = \frac{6}{L} E^{7 8 9 \natural},
\end{eqnarray}
where we abbreviated $E^{a_{1}} \w \cdots \w E^{a_{p}}$ as $E^{a_{1} \cdots a_{p}}$. In the following sections, all indices of field variables represent the ones in the local Lorentz frame.


\subsection{M5-brane wrapping $\AdS_{3} \times S^{3}$}
Here we consider an M5-brane wrapping $\AdS_3\times S^3$. In this calculation, we should carefully introduce the boundary term of the M5-brane action. Let us first consider the boundary term in the plane Wilson surface in $\mathbb{R}^6$ for simplicity. It is convenient to introduce the Poincar\'e coordinates
\begin{align}
d s^{2} =& \frac{L^{2}}{y^{2}} \lp d y^{2} + d r_{1}^{2} + r_{1}^{2} d \phi^{2} + d r_{2}^{2} + r_{2}^{2} d \Om_{3}^{2} \rp + \frac{L^{2}}{4} d \Om_{4}^{2},\nonumber\\
& y>0,\quad r_1,r_2\ge 0.
 \label{Poincare}
\end{align}
The plane Wilson surface is located at $r_2=0, y\to 0$.  We denote one of the worldvolume coordinates on the M5-brane by $\la$, and take the ansatz
\begin{eqnarray} 
r_{2} = \kappa y, \hspace{2em}
y = y ( \la ),
\end{eqnarray}
where $\kappa$ is a constant. The induced metric is given by
\begin{eqnarray} 
	\begin{aligned}
	d s_{\rm ind}^{2} &= \frac{L^{2}}{y^{2}} \lb \lp 1 + \kappa^{2} \rp y'^2 d \la^{2} + d r_{1}^{2} + r_{1}^{2} d \phi^{2} + ( \kappa y )^{2} d \Om_{3}^{2} \rb, \\ 
	\sqrt{g_{\rm ind}} &= \frac{\kappa^{3} L^{6}}{y^{3}} | y' | r_{1} \sqrt{1 + \kappa^{2}} \sqrt{g_{S^{3}}}, 
	\end{aligned}
\end{eqnarray}
where $y':=dy/d\lambda$, and $g_{S^3}$ is the determinant of the metric of unit $S^3$.

Since the submanifold totally belongs to $\AdS_{7}$, we take account of the 7-form field strength $B_{7}$ which is the Hodge dual to $B_{4}$,
{\set
\begin{eqnarray} 
B_{7}
&=& \ast B_{4} \non \\ 
&=& \frac{6}{L} E^{0 1 2 3 4 5 6} \non \\ 
&=& \frac{6 L^{6}}{y^{7}} r_{1} r_{2}^{3} d y \w d r_{1} \w d \phi \w d r_{2} \w \om_{3},
\end{eqnarray}}%
where $\om_3$ is the volume form of unit $S^3$. $B_{7}$ can be written as the following form with background gauge fields $C_{3}$ and $C_{6}$ to satisfy the equation of motion for $B_{4}$:
\begin{eqnarray}
B_{7} = d C_{6} + \frac{1}{2} C_{3} \w d C_{3}.
\end{eqnarray}
Since $C_{3} \w d C_{3} = 0$, we choose the gauge in which $C_{6}$ is given by
{\set
\begin{eqnarray} 
C_{6}
&=& - \frac{L^{6}}{y^{6}} r_{1} r_{2}^{3} d r_{1} \w d \phi \w d r_{2} \w \om_{3} \non \\ 
&=&  \frac{\kappa^{4} L^{6}}{y^{3}} r_{1}  y'  d r_{1} \w d \phi \w \om_{3}\w d\la. \label{c6} 
\end{eqnarray} }%
There is the 2-form gauge field $A_{2}$ on the M5-brane and let us define $F_3=dA_2$ and $H_3=F_3-C_3$.  Notice that $C_3=0$ on this M5-brane worldvolume. The flux quantization condition \eqref{fq} implies
{\set
\begin{eqnarray} 
H_{3}
&=& \frac{k}{2 N} L^3\omega_3 =\frac{k}{2 N} \frac{y^{3}}{r_{2}^{3}} E^{4 5 6} \non \\[.5em] 
\RA H_{4 5 6}
&=& \frac{k}{2 N \kappa^{3}}. 
\end{eqnarray} }%
We use the gauge symmetry \eqref{gauge2} and set 
\begin{align}
H_{012}=0. \label{H012}
\end{align}
Actually, the final result is independent of this gauge choice as far as we use the Legendre transformation prescription for the 2-form gauge field as in \cite{Drukker:1999zq,Drukker:2005kx}. In order to determine the field strength $\tilde{H}_{3}$ dual to $H_{3}$, we must fix an auxiliary field $a$ which makes the action covariant (see Appendix \ref{PSTaction}). Through the rest of this paper, we use
\begin{eqnarray} 
a = \phi \label{afix},
\end{eqnarray}
then
\begin{eqnarray} 
v_{2} = 1.
\end{eqnarray}

The component of $\tilde{H}^{a b}$ left under the fixing \eqref{afix} is
\begin{eqnarray} 
\tilde{H}^{0 1} = H_{4 5 6}. \label{dh3}
\end{eqnarray}
Since the PST action \eqref{PST} is originally defined in the Lorentzian background, we make the Wick rotation $\tilde{H}_{t 1} = i \tilde{H}_{\tau 1}$. Accordingly, the PST action \eqref{PST} with nonzero $C_{6}$ becomes
{\set
\begin{eqnarray} 
S_{\rm M5}
&=& T_{5} \int d^{6} \zeta \sqrt{g_{\rm ind}} \sqrt{\det \lp \dl_{m}^{\ n} + i \tilde{H}_{m}^{\ n} \rp}
+ T_{5} \int C_{6} \non \\[.5em] 
&=& \Kcal \int_{\la_{\rm min}}^{\la_{0}} d \la \frac{| y' |}{y^{3}} \lb \sqrt{\lp 1 + \kappa^{2} \rp \lp \kappa^{6} + \left( \frac{k}{2 N} \right)^{2} \rp}
- \kappa^{4} \rb, 
\end{eqnarray} }where
\begin{eqnarray}
\Kcal := 2 \pi^{2} T_{5} L^{6} \int_{0}^{2 \pi} d \phi \int_{0}^{\infty} d r_{1} r_{1}.
\end{eqnarray}
We assume $y'<0$ and introduce the cutoff denoted by $\la_{0}$ and the lower bound $\la_{\rm min}$.   The equation of motion for $\kappa$ is
{\set
\begin{eqnarray} 
0
&=& \frac{d}{d \kappa} \lb \sqrt{\lp 1 + \kappa^{2} \rp \lp \kappa^{6} + \left( \frac{k}{2 N} \right)^{2} \rp}
- \kappa^{4} \rb \non \\[.5em] 
&=& \frac{\left( \frac{k}{2 N} \right)^{2} \kappa + 3 \kappa^{5} + 4 \kappa^{7}}{\sqrt{\lp 1 + \kappa^{2} \rp \lp \kappa^{6} + \left( \frac{k}{2 N} \right)^{2} \rp}}
- 4 \kappa^{3}, 
\end{eqnarray} }%
hence $\kappa$ is related to $k$ by
\begin{eqnarray}
\kappa = \sqrt{\frac{k}{2 N}}. 
\end{eqnarray}
We can rewrite the action with this relation as
\begin{eqnarray} 
S_{\rm M5} = \Kcal \frac{k}{2 N} \int_{\la_{\rm min}}^{\la_{0}} d \la \frac{| y' |}{y^{3}}.
\end{eqnarray}
Furthermore, we replace the bulk direction $y$ with $z$ such that
\begin{eqnarray} 
z = \frac{1}{y^{2}}.
\end{eqnarray}
Because $z ( \la_{\rm min} ) = 0$ in the new coordinate, the PST action is given by
{\set
\begin{eqnarray} 
S_{\rm M5}
&=& \frac{k}{4 N} \Kcal \int_{\la_{\rm min}}^{\la_{0}} d \la z' = : \int_{\la_{\rm min}}^{\la_{0}} d \la \Lcal \non \\[.5em] 
&=& \frac{k}{4 N} \Kcal z_{0}, \label{ppst} 
\end{eqnarray} }where a new cutoff is defined as $z_{0} := z ( \la_{0} )$. Along the procedure of the Legendre transformation, we should impose the boundary condition on the conjugate momentum $P_{z}$ for $z$.\footnote{The coordinate $z$ is identified, up to a constant factor, with the radial coordinate of the asymptotically flat supergravity solution of M5-branes before taking the near horizon limit. Thus, this Legendre transformation is the analogue of the case of the Wilson loop case \cite{Drukker:1999zq,Drukker:2005kx}.}
 We would like to set the condition where the variation of $P_{z}$ is zero on the boundary,
\begin{eqnarray}
\dl P_{z} |_{\rm bdy} = 0.
\end{eqnarray}
The conjugate momentum is given by
\begin{eqnarray}
P_{z} = \frac{\p \Lcal}{\p z'} = \frac{k}{4 N} \Kcal,
\end{eqnarray}
and the boundary term can be written as
\begin{eqnarray} 
S_{\rm bdy} = - P_{z} z_{0}. \label{boundary-term}
\end{eqnarray}
We bring it and the original action together. Then the regularized action $S_{\rm M5}^{\rm reg}$ becomes
\begin{eqnarray}
S_{\rm M5}^{\rm reg} = S_{\rm M5} + S_{\rm bdy} = 0.
\end{eqnarray}
Thus, the expectation value for the M5-brane is 1. This result is expected since the plane Wilson surface preserves a part of the Poincar\'e supersymmetry. The boundary term \eqref{boundary-term} is proportional to the volume of the boundary including the finite contribution. Thus, we conclude that the boundary counter term  is proportional to the volume of the boundary with the gauge choice \eqref{c6} and \eqref{H012}.

Let us move to the Wilson surface wrapping on $S^1\times S^1$.  It is convenient to use the $\AdS_3\times S^3$ foliation coordinates \eqref{bubbling} with identification \eqref{identification}.
They are related by the coordinate transformation:
\begin{eqnarray} 
	\begin{aligned}
	y &= \frac{e^{\tau}}{\cosh u \cosh w}, \\[.5em] 
	r_{1} &= e^{\tau} \tanh w, \\[.5em] 
	r_{2} &= \frac{e^{\tau} \tanh u}{\cosh w}. 
	\end{aligned} \label{PtoGlobal}
\end{eqnarray}
The M5-brane is wrapping $\AdS_3\times S^3$ expressed by $u=u_k=$(constant). From \eqref{PtoGlobal}, $\kappa$ is related to $u_k$ as
\begin{eqnarray} 
\kappa = \sinh u_{k}.
\end{eqnarray}
Similary, $C_{6}$ on the worldvolume is given by
\begin{eqnarray}
C_{6} = - L^{6} \cosh^{2} u_{k} \sinh^{4} u_{k} \cosh w \sinh w d \tau \w d w \w d \phi \w \om_{3}.
\end{eqnarray}
In addition, we must use the flux quantization condition in this coordinate, namely, $H_{3}$ is given by
{\set
\begin{eqnarray}
H_{3}
&=& \frac{k}{2 N \sinh^{3} u_{k}} E^{4 5 6} \non \\ 
\RA H_{4 5 6}
&=& \frac{k}{2 N \sinh^{3} u_{k}}. 
\end{eqnarray} }On the other hand, \eqref{dh3} remains intact. Putting it all together, we can compute the PST action in these coordinates,
{\set
\begin{eqnarray}
S_{\rm M5}
&=& T_{5} \int L^{6} \om_{6} \cosh^{3} u_{k} \sinh^{3} u_{k} \sqrt{1 + \lp H_{4 5 6} \rp^{2}} \non \\[.5em] 
&& - T_{5} L^{6} \int \cosh^{2} u_{k} \sinh^{4} u_{k} \cosh w \sinh w d \tau \w d w \w d \phi \w \om_{3} \non \\[.5em] 
&=& \frac{2 \pi R_{6}}{r} k \lp 2 N + k \rp \sinh^{2} w_{0} ,
\end{eqnarray} }where $\om_{6}$ is the volume form of unit $\AdS_{3} \times S^{3}$ and $w_{0}$ is a cutoff. 
Since the boundary term is proportional to the volume of the boundary and cancels the divergence,
it is given by
\begin{align}
S_{\rm bdy}=-\frac{2 \pi R_{6}}{r} k \lp 2 N + k \rp \sinh w_{0} \cosh w_0.
\end{align}
The regularized PST action $S_{\rm M5}^{\rm reg}$ is obtained in the limit $w_0\to \infty$ as
{\set
\begin{eqnarray}
S_{\rm M5}^{\rm reg}
&=& S_{\rm M5} + S_{\rm bdy} \non \\[.5em] 
&=& - \frac{\pi R_{6}}{r} k \lp 2 N + k \rp \non \\[.5em] 
&=& - \frac{\beta}{2} N k \lp 1 + \frac{k}{2N} \rp. 
\end{eqnarray} }Finally, the expectation value of the Wilson surface for the M5-brane wrapping $\AdS_{3} \times S^{3}$ is given by
\begin{eqnarray} 
\exp \lb - S_{\rm M5}^{\rm reg} \rb = \exp \lb \frac{\beta}{2} N k \lp 1 + \frac{k}{2N} \rp \rb.
\end{eqnarray}
This result completely matches the value of the Wilson surface in symmetric representation \eqref{WLs}. As a result, we could obtain nontrivial support for the $\AdS_{7}$/CFT$_{6}$.

\subsection{M5-brane wrapping $\AdS_{3} \times \tilde{S}^{3}$}
In this section we consider a probe M5-brane wrapping $\AdS_{3} \times \tilde{S}^{3}$.  Here $\AdS_{3}$ is a minimal surface in $\AdS_{7}$,  while $\tilde{S}^{3}$ is included in $S^{4}$. It is convenient to use the global coordinates \eqref{global}.  We take the ansatz
\begin{align}
\eta=\pi/2,\quad \theta=\theta_k=(\text{constant}).
\end{align}
The induced metric on the M5-brane is given by
\begin{eqnarray}
	\begin{aligned}
	d s_{\rm ind}^{2} &= L^{2} \lp \cosh^{2} \rho d \tau^{2} + d \rho^{2} + \sinh^{2} \rho d \phi^{2} \rp + \frac{L^{2}}{4} \sin^{2} \theta_{k} d \tOm_{3}^{2}, \\ 
	\sqrt{g_{\rm ind}} &= \frac{L^{6}}{8} \cosh \rho \sinh \rho \sin^{3} \theta_{k} \sqrt{g_{\tilde{S}^{3}}}, 
	\end{aligned} \label{induced1}
\end{eqnarray}
where constant $\theta_{k}$ is associated with integer $k$ parametrizing the flux quantization condition (see Appendix \ref{Flux}).

$B_{4}$ also can be written as the derivative of $C_{3}$; thus, for the global coordinates we have
{\set
\begin{eqnarray}
B_{4} = d C_{3}
&=& \frac{6}{L} E^{7 8 9 \natural} \non \\[.5em] 
&=& \frac{3}{8} L^{3} \sin^{3} \theta d \theta \w \tom_{3},
\end{eqnarray} }where $\tom_{3}$ is the volume form of unit $\tilde{S}^{3}$. By integrating this over $\theta$, $C_{3}$ can be obtained by
{\set
\begin{eqnarray}
C_{3}
&=& - \frac{L^{3}}{8} \lp 3 \cos \theta - \cos^{3} \theta - 2 \rp \tom_{3} \\[.5em]
&=:& - L^{3} f ( \theta ) \tom_{3}.
\end{eqnarray} }
We choose the gauge in which $C_{3} = 0$ at $\theta = 0$ because $\tilde{S}^{3}$ shrinks at that point. Combining it with the flux quantization condition \eqref{fq}, the 3-form field strength $H_{3}$ is
{\set
\begin{eqnarray} 
H_{3}
&=& F_{3} - C_{3} \non \\[.5em] 
&=& \lp \frac{k}{2 N} + f ( \theta_{k} ) \rp L^{3} \tom_{3} \non \\[.5em] 
&=& \lp \frac{k}{2 N} + f ( \theta_{k} ) \rp \frac{8}{\sin^{3} \theta_{k}} E^{8 9 \natural} \non \\[.5em] 
\RA H_{8 9 \natural}
&=& \lp \frac{k}{2 N} + f ( \theta_{k} ) \rp \frac{8}{\sin^{3} \theta_{k}}. \label{H1} 
\end{eqnarray} } The component of $\tilde{H}^{a b}$ is
\begin{eqnarray} 
\tilde{H}^{0 1} = H_{8 9 \natural}. \label{dualH1}
\end{eqnarray}

We choose the gauge $H_{012}=0$ again. Moreover, we have $C_{6}= 0$ on the worldvolume and $C_{3} \w H_{3} = 0$ because both are proportional to the volume form of $\tilde{S}^{3}$. Thus, the remaining part of the action is
{\set
\begin{eqnarray} 
S_{\rm M5}
&=& T_{5} \int d^{6} \zeta \sqrt{g_{\rm ind}} \sqrt{\det \lp \dl_{m}^{\ n} + i \tilde{H}_{m}^{\ n} \rp} \non \\[.5em] 
&=& T_{5} \int d^{6} \zeta \frac{L^{6}}{8} \cosh \rho \sinh \rho \sin^{3} \theta_{k} \sqrt{g_{\tilde{S}^{3}}} \sqrt{1 + \lp H_{8 9 \natural} \rp^{2}} \non \\[.5em] 
&=& T_{5} \frac{\pi^{2} L^{6}}{4} \int d^{3} \zeta \cosh \rho \sinh \rho \sqrt{\sin^{6} \theta_{k} + 64 \lp \frac{k}{2 N} + f ( \theta_{k} ) \rp^{2}}. 
\end{eqnarray} }%
Next we solve the equation of motion for $\theta_{k}$ to acquire the on-shell value. It is equivalent to
{\set
\begin{eqnarray} 
0
&=& \frac{d}{d \theta_{k}} \lb \sin^{6} \theta_{k} + 64 \lp \frac{k}{2 N} + f ( \theta_{k} ) \rp^{2} \rb \non \\[.5em] 
&=& 8 \sin^{3} \theta_{k} \lb - 2 \cos \theta_{k} - \frac{4 k}{N} + 2 \rb, 
\end{eqnarray} }then we have the relation between $\theta_{k}$ and $k$ as
\begin{eqnarray} 
\cos \theta_{k} = 1 - \frac{2 k}{N}.
\end{eqnarray}
Substituting this into the action, we obtain
{\set
\begin{eqnarray} 
S_{\rm M5}
&=& T_{5} \frac{\pi^{2} L^{6}}{4} \frac{4 k}{N} \lp 1 - \frac{k}{N} \rp \int_{0}^{\rho_{0}} d \rho \int_{0}^{\frac{2 \pi R_{6}}{r}} d \tau \int_{0}^{2 \pi} d \phi \cosh \rho \sinh \rho \non \\[.5em] 
&=& \frac{4 \pi R_{6}}{r} k ( N - k ) \sinh^{2} \rho_{0}, 
\end{eqnarray} }where $\rho_{0}$ is a cutoff.
The boundary term $S_{\rm bdy}$ is again proportional to the volume of the boundary and given by
\begin{align}
S_{\rm bdy}=-\frac{4 \pi R_{6}}{r} k ( N - k ) \sinh \rho_{0} \cosh \rho_0.
\end{align}
We take the limit $\rho_{0} \ra \infty$,  and obtain  {\set
\begin{eqnarray} 
S_{\rm M5}^{\rm reg}
&=& S_{\rm M5} + S_{\rm bdy} \non \\[.5em] 
&=& - \frac{2 \pi R_{6}}{r} k ( N - k ) \non \\[.5em] 
&=& - \frac{\beta}{2} Nk \left( 1 - \frac{k}{N} \right). 
\end{eqnarray} }The expectation value for the M5-brane wrapping $\AdS_{3} \times \tilde{S}^{3}$ results in
\begin{eqnarray} 
\exp \lb - S_{\rm M5}^{\rm reg} \rb = \exp \lb \frac{\beta}{2} Nk \left( 1 - \frac{k}{N} \right) \rb.
\end{eqnarray}
It perfectly agrees with the Wilson surface in anti-symmetric representation \eqref{WLa}; hence, this strongly stands for the $\AdS_{7}$/CFT$_{6}$.

\subsection*{Acknowledgements}
We would like to thank Yuhma Asano, Koji Hashimoto, Masazumi Honda, Kazuo Hosomichi, Yosuke Imamura, Hiroshi Isono, Johan K\"all\'en, Yoichi Kazama, Shota Komatsu, Sanefumi Moriyama, Tomoki Nosaka, Takuya Okuda, Yuji Okawa, Akinori Tanaka, and Seiji Terashima for discussions and comments. The work of H.M. was supported in part by the JSPS Research Fellowship for Young Scientists. The work of S.Y. was supported in part by JSPS KAKENHI Grant No. 22740165.

\appendix
\section{Calculation of the Wilson surface in a rectangular Young diagram}
Here we calculate the expectation value of the Wilson surface in a rectangular Young diagram following the formulation of Halmagyi and Okuda \cite{Halmagyi:2007rw}.  Let the height of the rectangular Young diagram be $n$ and the width $m$. Then it has been found that the Wilson loop expectation value in the Chern-Simons matrix model is expressed as
\footnote{The notation of the integration variables here is related to \cite{Halmagyi:2007rw} by $u^{(1)}_{i}=N \nu_i,\ (i=1,\dots,n)$ and $u^{(2)}_{a-n}=N \nu_{a},\ (a=n+1,\dots,N)$.}
\begin{align}
\WLR_{{\rm U}(N)}=\frac{1}{\Zcal}\int \prod_{A=1}^{N}d\nu_A  \exp \Fcal,
\label{U(N)}
\end{align}
where $\Zcal$ is the appropriate normalization and $\Fcal$ is given by
\begin{align}
\Fcal:=&
-\frac{1}{\beta}N^2\sum_{i=1}^{n}\nu_i^2+\left(m+\frac12(N-n)\right)N\sum_{i=1}^n\nu_i
+\sum_{i,j,i<j} \ln\left|\sinh \frac{N}{2}(\nu_i-\nu_j)\right|^2\nonumber\\
&-\frac{1}{\beta}N^2\sum_{a=n+1}^{N}\nu_a^2+\left(-\frac12n\right)N\sum_{a=n+1}^N\nu_a
+\sum_{a,b,a<b} \ln\left|\sinh \frac{N}{2}(\nu_a-\nu_b)\right|^2\nonumber\\
&+\sum_{i=1}^{n}\sum_{a=n+1}^{N}\ln\left|\sinh \frac{N}{2}(\nu_i-\nu_a)\right|. \label{Fcal}
\end{align}
We would like to evaluate this integral in the limit $N,n,m\to \infty$ while $n/N, m/N$ are kept finite.  In this limit we can use the saddle point approximation. Eq. \eqref{Fcal} is simplified as
\begin{align}
\Fcal=&
-\frac{1}{\beta}N^2\sum_{i=1}^{n}\nu_i^2+\left(m+\frac12(N-n)\right)N\sum_{i=1}^n\nu_i
+N\sum_{i,j,i<j}\left|\nu_i-\nu_j\right|\nonumber\\
&-\frac{1}{\beta}N^2\sum_{a=n+1}^{N}\nu_a^2+\left(-\frac12n\right)N\sum_{a=n+1}^N\nu_a
+N\sum_{a,b,a<b} \left|\nu_a-\nu_b\right|\nonumber\\
&+\frac{N}{2}\sum_{i=1}^{n}\sum_{a=n+1}^{N}\left|\nu_i-\nu_a\right|. \label{Fcal2}
\end{align}
The saddle point equations are derived from eq \eqref{Fcal2} as
\begin{align}
&-\frac{2}{\beta} N^2 \nu_i + \left(m+\frac12(N-n)\right)N+N\sum_{j,j\ne i}\sign(\nu_i-\nu_j)
+\frac{N}{2}(N-n)=0,\quad i=1,\dots,n,\nonumber\\
&-\frac{2}{\beta} N^2 \nu_a -\frac12 nN+N\sum_{b,b\ne a}\sign(\nu_a-\nu_b)
-\frac{N}{2}n=0,\quad a=n+1,\dots,N.
\end{align}
If we assume the order
\begin{align}
\nu_A >\nu_B,\quad \text{if} \ A<B, \quad (A,B=1,\dots,N),
\end{align}
the solution is given by
\begin{align}
\nu_i=\frac{\beta}{2N}(m+N-2i),\quad(i=1,\dots,n),\nonumber\\
\nu_a=\frac{\beta}{2N}(N-2a),\quad(a=n+1,\dots,N).
\label{Saddle}
\end{align}
In other words the eigenvalue density is expressed as
\begin{align}
\rho(\nu)=
\begin{cases}
\frac{1}{\beta},& (-\frac{\beta}{2} <\nu<\frac{\beta}{2N}(N-2n),\ 
\frac{\beta}{2N}(N+m-2n) < \nu <\frac{\beta}{2N}(N+m)),\\
0 & (\text{others}).
\end{cases}
\end{align}
This is a special case of the eigenvalue distribution obtained in \cite{Halmagyi:2007rw}.\footnote{The results are the same although they first take the 't Hooft limit and then take the strong coupling limit in \cite{Halmagyi:2007rw}.} The expectation value \eqref{U(N)} is evaluated as
\begin{align}
\WLR_{{\rm U}(N)} \sim & \exp \Fcal |_{\text{saddle point}}\nonumber\\
 \sim & \exp \left[
 \frac{\beta}{2}mnN\left(1-\frac{n}{N}+\frac{m}{2N}\right)
 \right].\label{WRect}
\end{align}
This equation reproduces the result of the symmetric representation \eqref{WLs} when $n=1,\ m=k$, and the anti-symmetric representation \eqref{WLa} when $n=k,m=1$.

The result \eqref{WRect} is not invariant under the exchange of $n$ and $(N-n)$ because this is the expectation value in the U$(N)$ theory. It is related to the SU$(N)$ theory by
\begin{align}
\WLR_{{\rm U}(N)}=e^{\frac{\beta |R|^2}{4N}} \WLR_{{\rm SU}(N)},
\end{align}
where $|R|$ is the number of boxes in the Young diagram $R$. We obtain the expectation value in the SU$(N)$ theory by making use of this relation as
\begin{align}
\WLR_{{\rm SU}(N)}\sim \exp\left[
\frac{\beta}{2}mnN\left(1-\frac{n}{N}\right)\left(1+\frac{m}{2N}\right)
\right].\label{WRectSU}
\end{align}
This is invariant under the exchange of $n$ and $(N-n)$ as expected.  This expectation value also reproduces the results \eqref{WLs} and \eqref{WLa}.

\section{Flux quantization condition} \label{Flux}
We explain the flux quantization for the coupling of a probe M5-brane involving $S^{3}$ to an open M2-brane electrically following \cite{Camino:2001at}. We denote the worldvolume manifold of the M2-brane by $\Sg_{3}$ whose boundary $\p \Sg_{3}$ is part of the worldvolume of the M5-brane. For simplicity, $\p \Sg_{3}$ is the boundary of a disk $D^{3}$ embedded into the M5-brane.  Moreover, $\Sg_{4}$ represents the four-manifold with boundaries $\Sg_{3}$ and $D^{3}$. If we consider the coupling of the M5-brane and the M2-brane, the interaction term is written as
\begin{eqnarray}
S_{\rm int} [ \Sg_{4}, D^{3} ] = T_{2} \int_{\Sg_{4}} B_{4} + T_{2} \int_{D^{3}} H_{3}, 
\end{eqnarray}
where $T_{2} = \frac{1}{( 2 \pi )^{2} \lpl^{3}}$ is the tension of the M2-brane. 

In general, the action itself depends on the choice of $(D^3,\Sigma_4)$, though the weight with it in the path integral should be independent of such choice. Let $(D^{3'},\Sigma'_4)$ be another choice and we require that
\begin{eqnarray}
e^{i S_{\rm int} [ \Sg_{4}, D^{3} ]} = e^{i S_{\rm int} [ \Sg'_{4}, D^{3'} ]}. \label{weight}
\end{eqnarray}
This gives us the quantization condition for the flux through $S^{3}$ wrapped by the M5-brane. The condition \eqref{weight} can be written as
{\set
\begin{eqnarray}
2 \pi k
&=& S_{\rm int} [ \Sg_{4}, D^{3} ] - S_{\rm int} [ \Sg'_{4}, D^{3'} ] \non \\[.5em] 
&=& T_{2} \int_{\Sg_{4} - \Sg'_{4}} B_{4} + T_{2} \int_{D^{3} - D^{3'}} H_{3} \non \\[.5em] 
&=& T_{2} \int_{\Bcal^{4}} d C_{3} + T_{2} \int_{S^{3}} \lp F_{3} - C_{3} \rp \non \\[.5em] 
&=& T_{2} \int_{S^{3}} F_{3}, 
\end{eqnarray} }where $k \in \mathbb{Z}$, $F_3=dA_2$, and $A_2$ is the worldvolume 2-form gauge field. Since $F_{3}$ is proportional to the volume form $\om_{3}$ of the unit $S^{3}$, we obtain the flux quantization condition
\begin{eqnarray} 
F_{3}
= \frac{k}{\pi T_{2}} \om_{3}
= \frac{k}{2 N} L^{3} \om_{3}. \label{fq}
\end{eqnarray}

\section{PST action} \label{PSTaction}
The PST action proposed by \cite{Pasti:1997gx, Bandos:1997ui, Bandos:1997gm} is the covariant action on a single M5-brane.  Let $\zeta^m (m=0,1,\dots,5)$ be the worldvolume coordinates. The bosonic fields contain a scalar field $a$ and a 2-form gauge field $A_2=\frac{1}{2}A_{mn}d\zeta^{m}\wedge d\zeta^{n}$ as well as the spacetime coordinates.  
The bosonic part of the action with the Wess-Zumino term is given by
\begin{eqnarray} 
S_{\rm M5} = T_{5} \int d^{6} \zeta \sqrt{- g_{\rm ind}} \lb \Lcal + \frac{1}{4} \tilde{H}^{m n} H_{m n} \rb + T_{5} \int \lp C_{6} - \frac{1}{2} C_{3} \w H_{3} \rp, \label{PST}
\end{eqnarray}
where
{\set
\begin{eqnarray} 
\Lcal &=& \sqrt{\det \lp \dl_{m}^{\ n} + i \tilde{H}_{m}^{\ n} \rp}, \\ 
F_3&=&dA_2,\\
H_{3} &=& F_{3} - C_{3}, \\ 
H_{m n} &=& H_{m n p} v^{p}, \\ 
\tilde{H}^{m n} &=& ( \ast_{6} H )^{m n p} v_{p}, \\ 
v_{p} &=& \frac{\p_{p} a}{\sqrt{- g^{m n} \p_{m} a  \p_{n} a }}. 
\end{eqnarray} }%
The indices are raised or lowered by the induced metric.  The Hodge star $\ast_{6}$ is defined with the induced metric on the M5-brane. In addition, the action is invariant under the gauge transformation $\dl_{g}$
\begin{eqnarray} 
\dl_{g} A_{m n} = \p_{[ m} \phi_{n ]} ( \zeta ),
\end{eqnarray}
and the following local transformations $\dl_{\varphi}$ and $\dl_{\psi}$:
\begin{eqnarray} 
	&& \lc 
	\begin{aligned}
	\dl_{\varphi} a &= 0, \\ 
	\dl_{\varphi} A_{m n} &= \frac{1}{2} \p_{[ m} a\ \varphi_{n ]} ( \zeta ), 
	\end{aligned} \right. \label{gauge2}\\[.5em]
	&& \lc 
	\begin{aligned}
	\dl_{\psi} a &= \psi ( \zeta ), \\ 
	\dl_{\psi} A_{m n} &= - \frac{\psi ( \zeta )}{2 \sqrt{- g^{p q} \p_{p} a \p_{q} a }} \lp H_{m n} - \Vcal_{m n} \rp, 
	\end{aligned} \right.
\end{eqnarray}
where $\varphi_{m} ( \zeta )$ and $\psi ( \zeta )$ are infinitesimal parameters for each transformation, and
\begin{eqnarray}
\Vcal_{m n} := - 2 \frac{\dl \Lcal}{\dl \tilde{H}^{m n}}.
\end{eqnarray}


\providecommand{\href}[2]{#2}\begingroup\raggedright\endgroup


\end{document}